# A Spectral Library and Method for Sparse Unmixing of Hyperspectral Images in Fluorescence Guided Resection of Brain Tumors

*David Black[1*], Benoit Liquet[2,3,4], Sadahiro Kaneko[5], Antonio Di Ieva[4,6], Walter Stummer[7], Eric Suero Molina[4,6,7*]*

[1]Department of Electrical and Computer Engineering, University of British Columbia, Vancouver, BC, Canada.
[2]School of Mathematical and Physical Sciences, Macquarie University, Sydney, Australia
[3]Laboratoire de Mathématiques et de ses Applications, E2S-UPPA, Université de Pau & Pays de L'Adour, France
[4]Computational NeuroSurgery (CNS) Lab, Macquarie University, Sydney, New South Wales, Australia
[5]Department of Neurosurgery, Hokkaido Medical Center, Hokkaido, Japan.
[6]Macquarie Medical School, Faculty of Medicine, Health and Human Sciences, Macquarie University, Sydney, New South Wales, Australia
[7]Department of Neurosurgery, University Hospital of Münster, Münster, Germany.

*Corresponding authors. E-mails: dgblack@ece.ubc.ca; eric.suero@ukmuenster.de;

***Running Head:*** *A Spectral Library for Unmixing in Hyperspectral Imaging*

## Abstract

Through spectral unmixing, hyperspectral imaging (HSI) in fluorescence-guided brain tumor surgery has enabled detection and classification of tumor regions invisible to the human eye. Prior unmixing work has focused on determining a minimal set of viable fluorophore spectra known to be present in the brain and effectively reconstructing human data without overfitting. With these endmembers, non-negative least squares regression (NNLS) was commonly used to compute the abundances. However, HSI images are heterogeneous, so one small set of endmember spectra may not fit all pixels well. Additionally, NNLS is the maximum likelihood estimator only if the measurement is normally distributed, and it does not enforce sparsity, which leads to overfitting and unphysical results. Here, we analyzed 555666 HSI fluorescence spectra from 891 ex vivo measurements of patients with various brain tumors to show that a Poisson distribution indeed models the measured data 82% better than a Gaussian in terms of the Kullback-Leibler divergence and that the endmember abundance vectors are sparse. With this knowledge, we introduce (1) a library of 9 endmember spectra, including PpIX (620 nm and 634 nm photostates), NADH, FAD, flavins, lipofuscin, melanin, elastin, and collagen, (2) a sparse, non-negative Poisson regression algorithm to perform physics-informed unmixing with this library without overfitting, and (3) a highly realistic spectral measurement simulation with known endmember abundances. The new unmixing method was then tested on the human and simulated data and compared to four other candidate methods. It outperforms previous methods with 25% lower error in the computed abundances on the simulated data than NNLS, lower reconstruction error on human data, better sparsity, and 31 times faster runtime than state-of-the-art Poisson regression. This method and library of endmember spectra can enable more accurate spectral unmixing to better aid the surgeon during brain tumor resection.

## Introduction

Delineating glioma margins during brain surgery is very difficult since the tumor is infiltrative and hard to distinguish from healthy tissue. However, fluorescence guidance can improve patient outcomes by increasing resection rates[1,2]. In fluorescence-guided resection (FGR) of brain tumors, the patient is given 20 mg/kg b.w. of 5-aminolevulinic acid (5-ALA) preoperatively. This gathers preferentially in tumor cells where it is metabolized to protoporphyrin IX (PpIX), a precursor on the heme synthesis pathway[3]. When excited with violet light at 405 nm, PpIX fluoresces red, with a double peak at 634 and 700 nm. The difference in wavelength between the excitation and emission is called the Stokes shift and allows the fluorescence to be isolated from the bright excitation light using optical filters[1,4]. Thus, tumors that are otherwise difficult to distinguish from healthy tissue can often be identified by their red glow. This leads to more complete resection and consequently better progression and overall survival[2,5]. However, in lower-grade glioma or infiltrating tumor margins, fluorescence is often not visible to humans. Using sensitive cameras does not improve the problem since the PpIX fluorescence is masked by other endogenous fluorophores known collectively as autofluorescence, which emit at similar wavelengths and intensities.

Hyperspectral imaging (HSI) allows the PpIX content to be isolated from autofluorescence by examining the emission spectrum. HSI devices capture three-dimensional data cubes containing all the scene's spectral and spatial information. Like an RGB image, which has three channels, data cubes can have hundreds of channels, each at a different wavelength. Each pixel, therefore, contains the full emission and reflectance spectrum of that point. Thus, the fluorescence spectra of every visible point in the image can be captured[6]. Each measured spectrum contains a combination of fluorescing molecules, or fluorophores, including PpIX and the aforementioned autofluorescence. A linear model is commonly assumed, in which the measured fluorescence spectrum ($\mathbf{s}$) is a linear combination of the emission spectra of the present fluorophores ($\mathbf{b}_i$), also called endmember spectra[7]:

$$\mathbf{s} = \sum_{i=1}^{k} c_i \mathbf{b}_i = B\mathbf{c} \qquad (1)$$

where $B = [\mathbf{b}_1 \cdots \mathbf{b}_k]$ is the endmember matrix. With prior knowledge of the endmember spectra, the relative abundances ($c_i$) of the endmembers can be estimated using linear regression techniques[8]. During 5-ALA-mediated fluorescence-guided surgery for malignant gliomas, the endmembers likely include the two photostates of PpIX[3,9], called $PpIX_{620}$ and $PpIX_{634}$, as well as autofluorescence from flavins, lipofuscin, NADH, FAD, melanin, collagen, and elastin[8,10]. However, only 3 or 4 endmembers are usually present in any given spectrum. This latter fact is called sparsity – the abundance vectors are sparsely populated with non-zero values.

The linear model neglects multiple scattering[11] and other nonlinear effects but gives a convenient, dimensionally-reduced representation of the spectra. It has been shown that almost all of the information of a given spectrum is contained in up to five endmember abundances[8,12]. As a result, recent work in HSI for fluorescence-guided surgery has shown great promise for detecting tumor regions[6,13] and classifying tissue types using the endmember abundances[12,14]. The abundances have also been used to study 5-ALA administration timing[15] and dosage[16] and to improve the image acquisition process itself[17–19]. However, these computations are very sensitive to the chosen endmember spectra and the unmixing method.

Many spectral unmixing methods have been proposed based on various regression[7,20–22], geometric[20], and deep learning[23–28] algorithms[29]. Nonlinear methods have also been explored[30–32]. Previous work in neurosurgery, however, has typically used non-negative least squares (NNLS) regression[3,8,12,13,33,34]. This is simple, fast, and enforces the physical constraint of abundance non-negativity. However, least squares is only the maximum likelihood estimate if the data is normally distributed. In fact, photon emission is theoretically Poisson-distributed[35]. Thus, others have proposed Poisson regression[36] methods. Without sparsity constraints, however, both of these methods overfit and thus improve in accuracy with the number of endmember spectra used; in reality, few fluorophores are present in a given pixel[7]. In this case, the output may fit the measurement extremely well, but it does not accurately describe the system's physical state. Not only is this undesirable, but it may also affect the accuracy of subsequent classification tasks performed with the abundances. Furthermore, it may attribute non-zero abundances to key fluorophores such as PpIX, potentially leading to false-positive tumor identification. Thus, various sparse methods have also been explored[7,20] through different norm regularizations, low-rank non-negative matrix factorization[37], and partial least squares[21,32,33].

To circumvent unmixing by obtaining semantic segmentations of tissue type directly from the raw data cube, various deep learning methods have been proposed[38–40]. For brain tumor resection, in particular, the technique is promising[41]. Studies have used random forests, support vector machines (SVMs), and convolutional neural networks (CNNs), or voting-based combinations of k-nearest neighbors (KNN), hierarchical k-means clustering, and data-driven dimensionality reduction techniques to perform segmentation of tissues in vivo[42–46]. Though promising and exciting, these papers obtained accuracies of around 70-80%[47], which is currently too low for clinical application, and were trained on relatively small datasets[48,49]. With commercial development[41] and further research, these values may increase, but the generalizability of such methods is questionable, especially across different devices or centers. Furthermore, they are fixed to performing a specific task, and their outputs are neither explainable nor guaranteed to fulfill any criteria of accuracy or robustness.

Therefore, a modular implementation of spectral correction followed by unmixing and further processing of the abundance vectors remains a flexible, generalizable, and robust methodology. For example, by adopting a modular approach, the output endmember abundances may be used to distinguish tumor from healthy tissue, classify the type of tumor, or analyze biomarkers such as isocitrate dehydrogenase (IDH) mutation, which is clinically highly relevant. Initial exploration of such processing has had promising results[12,14]. This classification module can be used on any device in any hospital, as long as the relevant endmember abundances are first computed. Similarly, since most patients should have predominantly the same fluorophores present, the unmixing can be completely general for any device. Only the preprocessing step is necessarily at least partly device-specific.

This paper, therefore, describes a practical, general, high-performance unmixing method and an associated library of endmember spectra for HSI in fluorescence-guided brain tumor resection. The method is fast, accurate, physics-informed, applicable to any device with requisite pre-processing, and is the maximum likelihood estimator for the unmixing. We first show that human brain HSI data is indeed Poisson distributed and sparse using a large and broadly diverse dataset including 184 patients and 891 fluorescence HSI data cubes. A method is then presented using this fact to produce highly realistic simulated data with known fluorophore abundances. Finally, the new unmixing method and four other candidate methods are applied to the real and simulated data in multiple experiments to compare their effectiveness. This paper thus elucidates the statistical nature of brain tumor HSI data and how to take advantage of its inherent structure through a complete unmixing method and library.

## **Methods**

**Imaging System**

The HSI device used in this paper has previously been described[6,8,12,13,15,16] and is shown in Fig. 1. Ex vivo tissue samples were illuminated in turn with blue light from a 405 nm LED, white light from an LED, and no light to capture fluorescence, white, and dark spectra respectively. During each illumination phase, the reflected light and emitted fluorescence were captured in a ZEISS Opmi Pico microscope (Carl Zeiss Meditec AG, Oberkochen, Germany) and passed through a liquid crystal tunable filter (Meadowlark Optics, Longmont, CO, USA) to a scientific metal oxide semiconductor (sCMOS) camera (PCO.Edge, Excelitas Technologies, Waltham, MA, USA). Hyperspectral data cubes were captured by sweeping the filter from 420 to 730 nm and capturing a 2048 x 2048 pixel grayscale image at every sampling wavelength. Each image had a 100 or 500 ms exposure time to ensure a good signal-to-noise ratio. Regions of 10 x 10 pixels were averaged to further increase the signal-to-noise ratio, and non-overlapping regions were extracted from each biopsy to ensure independent samples. The dark images were used to subtract the dark noise of the camera sensor. Next, the white reflectance spectra were used to correct the fluorescence spectra for geometric effects and inhomogeneous scattering and absorption across the surface using dual-band normalization[50,51]. Finally, the spectra were corrected for the filter transmission curves and wavelength-dependent sensitivity of the camera.

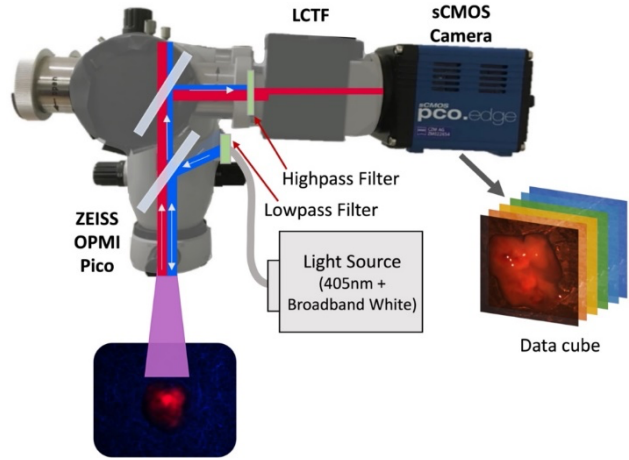

*Figure 1. Hyperspectral imaging device, with output data cube illustration.*

**Brain Tumor Data**

This HSI device has been used at the University Hospital of Münster, Münster, Germany, to examine ex vivo brain tumor samples removed during surgery and, as a result, obtain hyperspectral data cubes. A standard dose of 20 mg/kg of 5-ALA was administered orally to patients undergoing surgery for various brain tumors four hours before induction of anesthesia. Tissue resected by the surgeons was taken to the HSI system and imaged ex vivo before being passed on to pathology. Informed consent was obtained from each individual in this patient collective. All procedures performed in studies were in accordance with the ethical standards of the institutional and/or national research committee and with the 1964 Helsinki declaration and its later amendments or comparable ethical standards. Data collection and scientific use of biopsies had previously been approved by the ethics committee of the University of Münster.

| Class | # Samples | Class | # Samples |
|---|---|---|---|
| *Tissue Type* | *632* | *Margin Type (Gliomas)* | *288* |
| Pilocytic Astrocytoma | 5 | Reactive brain tissue | 100 |
| Diffuse Astrocytoma | 57 | Infiltrating zone | 57 |
| Anaplastic Astrocytoma | 51 | Solid tumor | 131 |
| Glioblastoma | 410 | | |
| Grade II Oligodendroglioma | 24 | *WHO Grade (Gliomas)* | *571* |
| Ganglioglioma | 4 | I | 9 |
| Medulloblastoma | 6 | II | 84 |
| Anaplastic Ependymoma | 8 | III | 57 |
| Anaplastic Oligodendroglioma | 4 | IV | 421 |
| Meningioma | 37 | | |
| Metastasis | 6 | *IDH Classification* | *411* |
| Radiation Necrosis | 20 | Mutant | 126 |
| | | Wildtype | 285 |

*Table 1. Overview of the evaluated dataset. In total, 891 hyperspectral data cubes were measured of ex vivo tissue from FGR of 184 patients.*

In total, data cubes of 891 biopsies from 184 patients were measured, resulting in 555666 human brain tumor spectra, described in Table 1. This large and diverse dataset was previously analyzed with reference to known fluorophores to determine the endmember spectra present in the data[8]. These included $PpIX_{634}$ and $PpIX_{620}$, the two fluorescing photostates of PpIX, as well as lipofuscin, flavins, and NADH. Fürtjes et al. similarly characterized five fluorophores, arriving at similar PpIX and lipofuscin spectra in addition to collagen, elastin, melanin, and FAD[10]. We combine these spectra into a single library of 9 endmembers, shown in Fig. 2. These are also available to download (see Supplementary Material).

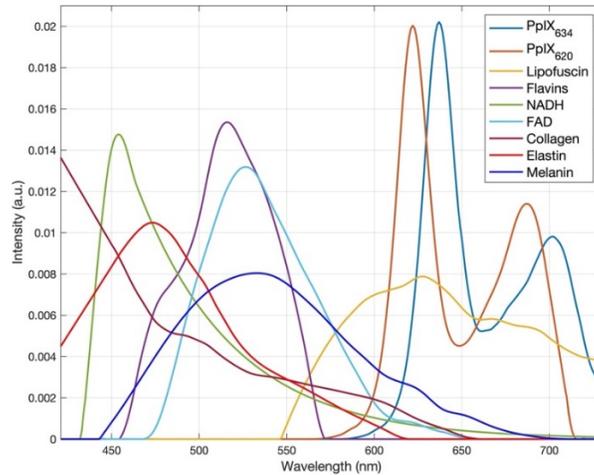

*Figure 2.* Library of nine endmember spectra to represent any brain tumor fluorescence HSI measurement. These include two PpIX and seven autofluorescence spectra.

While the measured spectra are assumed to be linear combinations of these endmembers, they additionally contain noise. Several measured spectra are shown in later figures. Assuming the measurements are normally distributed, the maximum likelihood estimate for the unmixing is the least squares solution[52]. However, in theory, the photon emission governing the measured spectra is Poisson distributed[35]. While normal distributions are described by their mean, μ, and variance, $\sigma^2$, Poisson distributions have a single parameter, $\lambda$, which equals both the mean and variance. Thus, regions of spectra with larger magnitude should also have more variance in the noise. If this is true, the maximum likelihood estimate for the unmixing would no longer be linear least squares. Therefore, we analyzed the data to determine its distribution.

To isolate the noise, it was not possible to unmix the spectra and then subtract the fitted spectrum from the measured one since the unmixing is imperfect. This leads to bias and strong artifacts and fails to isolate the noise. Instead, a high pass filter with relative frequency cut-off of 0.1 was used to remove the relatively low-frequency signal and keep only the noise. The cut-off frequency was chosen experimentally, as shown in Fig. 3.

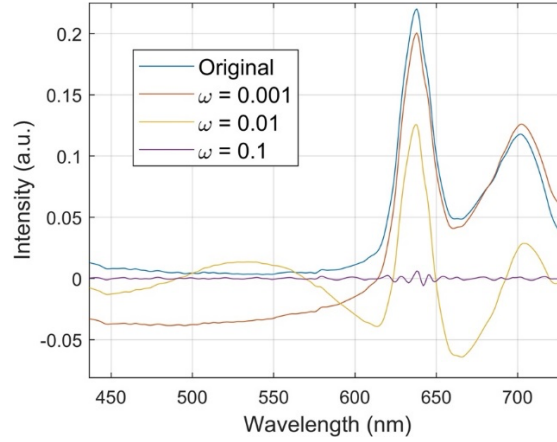

*Figure 3. Typical measured spectrum with high pass filters of various cut-off frequencies applied. A cut-off, $\omega = 0.1$ effectively isolates the noise by removing the main signal.*

This produces a distribution of 555666 noise magnitude values at each of the 310 sampling wavelengths. The mean and variance of these distributions were analyzed and correlated with the average magnitudes of the measured spectra at the corresponding wavelengths. Additionally, the parameter of a Poisson distribution describes the frequency of an event occurring. In this case, the event is the emission of photons, which occurs with an extremely high frequency even at relatively low light intensity, because each photon delivers such a small quantum of energy. Hence, $\lambda$ is very large, so the Poisson distribution is very closely approximated by a Gaussian with $\mu = \sigma^2 = \lambda$, where $\mu$ is the average magnitude of the measured spectra at the given wavelength. On the other hand, if the data is normally distributed, the variance should be independent of $\mu$. Hence, two Gaussian probability density functions (PDFs) were generated at each wavelength, both with mean $\mu$: one with variance $\sigma^2 = \mu$, and one with constant variance $\sigma^2 = v$. The constant $v$ was chosen as the average variance of all the measurement distributions. Each generated distribution was compared to the empirical distribution using the Kullback-Leibler (KL) divergence[53], which gives the level of difference between two distributions. The empirical distribution is not continuous, so the generated distribution was binned accordingly, and the discrete KL Divergence was used:

$$D_{KL}(p(x)||q(x)) = \sum_{x \in X} p(x) \ln \frac{p(x)}{q(x)}$$

**Simulation**

When evaluating an unmixing algorithm, simply comparing reconstruction error does not guarantee the underlying abundances are correct since many similarly effective unmixing solutions exist. Therefore, to assess the unmixing fully, it is necessary to have realistic data with known endmember abundances. With access to the noise model determined above, as well as the endmember spectra and statistics on their abundance distributions across 555666 human spectra, we can generate simulated spectra for this purpose that closely match real human data.

Assume the spectra are represented as $m \times 1$ vectors, I.E. they are sampled at $m$ wavelengths. Also, let $B \in \mathbb{R}^{m \times k}$ be the endmember matrix whose columns are the $k$ individual endmember spectra, $\boldsymbol{b}_i \in \mathbb{R}^m$. The simulated spectra were created as follows (the code is available in the Supplementary Materials):

1. The mean and variance of the distributions of the $k$ endmember abundances were extracted from the human data.
2. A set of $n$ artificial abundance vectors were sampled independently from normal distributions with these means and variances. The abundances form an abundance matrix $C_0 \in \mathbb{R}^{k \times n}$.
3. All abundances less than a threshold, $t$, were set to 0. In our arbitrary units, $t = 0.15$. This enforces the sparsity that is observed in human data.
4. The endmember matrix was multiplied by the abundance matrix to create a matrix of simulated spectra, $S \in \mathbb{R}^{m \times n}$: $S = BC_0$.
5. For each $m \times 1$ simulated spectrum, $\boldsymbol{s}$, a corresponding noise vector, $\boldsymbol{z}$, was generated such that each element $z_i$ was independently sampled randomly from a normal distribution with mean and variance equal to $s_i$.
6. The noise was added to the simulated spectrum, and the result was slightly smoothed using a Savitsky-Golay filter to simulate the smoothing from the image acquisition and interpolation process.

A set of n = 1000 resulting spectra is shown in Fig. 4 in comparison to a set of 1000 human measurement spectra. The two are virtually indistinguishable.

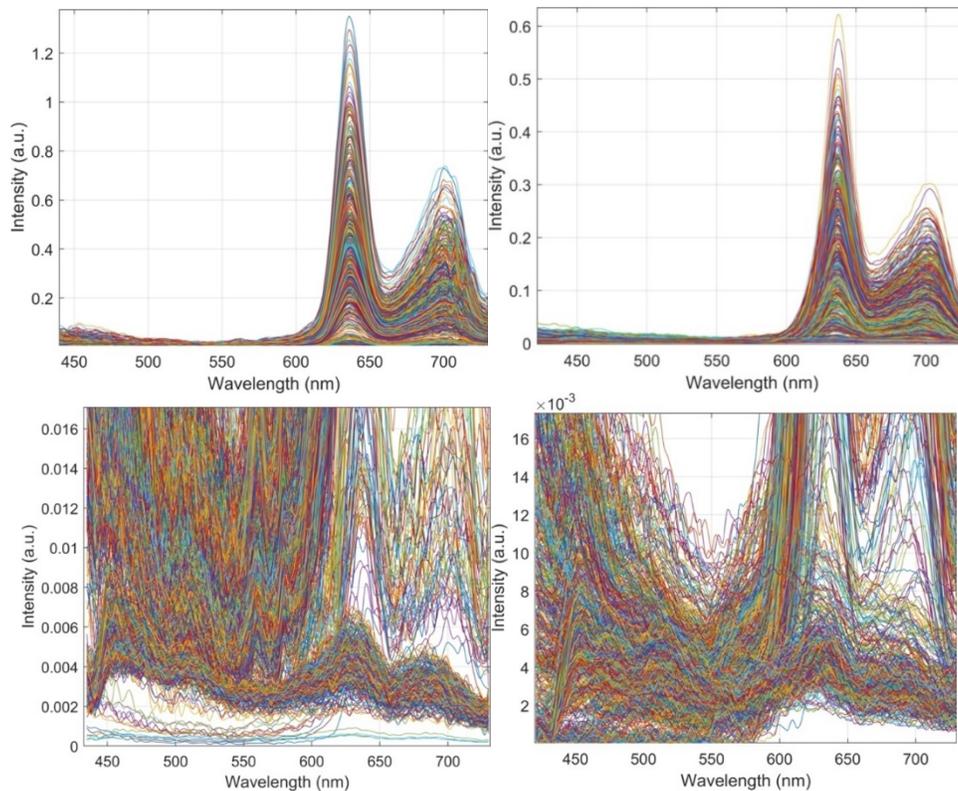

*Figure 4. 1000 real (left) and simulated (right) spectra, showing strong PpIX spectra (top), and zoomed-in weaker spectra containing predominantly autofluorescence (bottom). The real and simulated spectra of both sets are effectively indistinguishable.*

**Unmixing Methods**

For this paper, one new unmixing algorithm was developed, and several others were implemented for comparison. All implementations are found in the Supplementary Material, as are rigorous derivations of the algorithms. In this section, only the algorithms are outlined.

Two families of unmixing algorithms were tested, one using a least squares approach, in which the magnitude of the reconstruction error between the unmixing and the measured signal is minimized, and one using Poisson regression, in which the likelihood of the measurement given the abundance vector is maximized. The latter is maximum likelihood estimation (MLE), assuming Poisson-distributed measurements. Furthermore, in each algorithm except nonnegative least squares (NNLS, the baseline legacy method), the sparsity of the endmember abundances is enforced. The sparsity of a vector is given by its L0 norm, the number of non-zero elements in the vector. However, this is non-convex, so we use a relaxation in the form of the L1 norm: $\|x\|_1 = \sum |x_i|$, which is known as a Lasso model. In each case, we define a loss function $L(\boldsymbol{c}): \mathbb{R}_+^k \to \mathbb{R}$ and solve the following general optimization problem to obtain the estimate, $\hat{\boldsymbol{c}}$.

$$\hat{\boldsymbol{c}} = \underset{\boldsymbol{c} \geq 0}{\operatorname{argmin}} L(\boldsymbol{c}) \tag{2}$$

*Nonnegative Least Squares (NNLS)*

This is the method that has been previously used in brain tumor measurements. The objective is to minimize the L2 reconstruction error, and no regularization is used. The loss function is shown in Equation 3.

$$L(\boldsymbol{c}) = \frac{1}{2}\|\boldsymbol{s} - B\boldsymbol{c}\|^2 \tag{3}$$

The analytical solution to the unconstrained problem is the well-known least squares solution, $\hat{\boldsymbol{c}} = (B^\top B)^{-1} B^\top \boldsymbol{s}$. However, this does not account for the nonnegativity constraint. The constrained problem, however, is also solved very efficiently and exists as a built-in function in MATLAB (lsqnonneg.m), based on the algorithm by Lawson and Hanson[54].

*Sparse Nonnegative Least Squares (SNNLS)*

This simply extends NNLS with a sparsity constraint using L1 regularization, shown in Equation 4.

$$L(\boldsymbol{c}) = \frac{1}{2}\|\boldsymbol{s} - B\boldsymbol{c}\|^2 + \lambda \|\boldsymbol{c}\|_1 \tag{4}$$

As the L1 norm is non-differentiable at 0 and the problem is constrained, an analytical solution does not exist. However, due to the nonnegativity of $\boldsymbol{c}$, the L1 norm becomes $\mathbf{1}^\top \boldsymbol{c}$, so we can use projected gradient descent. The projection onto the feasible set is trivial: replace all negative elements with 0. We additionally use heavy-ball momentum[55], and an adaptive step size to improve the convergence. Hence, the problem is simply and efficiently solved using Algorithm 1.

**Algorithm 1** Heavyball Projected Gradient Descent for SNNLS
―――――――――――――――――――――――――――――――――
Given a measured spectrum, $s$, and a matrix of endmember spectra, $B$
Choose constants $\lambda = ?$ and $\mu = 0.9$
$c_0 \leftarrow \text{Proj}_{\mathbb{R}_+}\left((B^\top B)^{-1} B^\top s\right)$
$x_0 \leftarrow c_0$
**while** $\|c_t - c_{t-1}\|_2 > \epsilon$ **do**
  $\eta \leftarrow \frac{0.01}{\sqrt{t}}$
  $x_{t+1} \leftarrow \mu x_t + \lambda \mathbf{1}_k + B^\top(s - Bc)$
  $c_{t+1} \leftarrow \text{Proj}_{\mathbb{R}_+}(c_t - \eta x_{t+1})$
  $t \leftarrow t + 1$
**end while**
―――――――――――――――――――――――――――――――――

*Iterative Soft Thresholding Algorithm (ISTA)*
ISTA is a proximal gradient descent method to account for the non-smoothness of part of the objective function, as described by Beck and Teboulle[56] and shown in Algorithm 2. We use the same objective shown in Equation 4, which has a smooth and a non-smooth term. Due to this structure, the algorithm is much faster than subgradient methods. Accelerated versions of this, called fast ISTA (FISTA) exist[56], and were tested cursorily, but were excluded due to poor performance (see Discussion section).

**Algorithm 2** Iterative Soft Thresholding Algorithm (ISTA)
―――――――――――――――――――――――――――――――――
Given a measured spectrum, $s$, and a matrix of endmember spectra, $B$
Choose constant $\lambda = ?$
$c_0 \leftarrow \text{Proj}_{\mathbb{R}_+}\left((B^\top B)^{-1} B^\top s\right)$
**while** $\|c_t - c_{t-1}\|_2 > \epsilon$ **do**
  $\eta \leftarrow \frac{0.01}{\sqrt{t}}$
  $x \leftarrow c_t - 2\eta(B^\top(Bc_t - s) + \lambda \mathbf{1}_k)$
  $c_{t+1} \leftarrow \text{Proj}_{\mathbb{R}_+}((|x| - \lambda\eta \mathbf{1}_k) \odot \text{sign}(x))$
  $t \leftarrow t + 1$
**end while**
―――――――――――――――――――――――――――――――――

*Sparse Low-rank Poisson Regression (SLPR)*
This method and the next follow a completely different approach to the previous two methods, using Poisson regression through MLE, as described above. SLPRU was described by Wang et al.[36], and we used their MATLAB code. This method enforces not only sparsity in the abundance vector, but also low-rankness in the spatial distribution of the abundances. The latter decreases noise and creates a smoother overlay map of abundances. However, our preprocessing involves pixel averaging across a region of interest, so this is potentially redundant. Analysis of the necessity and comparison of the different methods for achieving spatial smoothness is left for future work.

*Sparse Nonnegative Poisson Regression (SNPR)*
This method is developed here as a much-simplified version of SLPR, which is more efficient and does not make the low-rank assumption. The objective function is shown in Equation 5, but the algorithm is derived in detail in the Supplementary Material.

$$L(c) = \mathbf{1}_m^\top Bc - s^\top \log Bc + \lambda \mathbf{1}_k^\top c \tag{5}$$

To solve this, we adopt a similar approach to the Algorithm 1, using projected gradient descent with heavy-ball momentum and adaptive step size. The projection operator is, again, trivial.

---

**Algorithm 3** Sparse Nonnegative Poisson Regression (SNPR)

Given a measured spectrum, $s$, and a matrix of endmember spectra, $B$
Choose constants $\lambda = ?$ and $\mu = 0.9$
$c_0 \leftarrow \text{Proj}_{\mathbb{R}_+}\left((B^\top B)^{-1} B^\top s\right)$
$x_0 \leftarrow c_0$
**while** $\|c_t - c_{t-1}\|_2 > \epsilon$ **do**
$\quad \eta \leftarrow \frac{0.01}{\sqrt{t}}$
$\quad x_{t+1} \leftarrow \mu x_t + \lambda \mathbf{1}_k + B^\top \mathbf{1}_m - \sum_{i=1}^m d_i \left(\frac{s_i}{d_i^\top c_t}\right)$
$\quad c_{t+1} \leftarrow \text{Proj}_{\mathbb{R}_+}(c_t - \eta x_{t+1})$
$\quad t \leftarrow t + 1$
**end while**

---

**Tests**
Several tests were performed to evaluate the unmixing methods. First, all the human data was unmixed using NNLS with different numbers of endmembers to examine how it overfits. Next, 1000 simulated spectra were unmixed using the different methods, and the reconstruction error, endmember abundance error, spectral angle error, runtime, and false positive rate were analyzed. The false positive rate is the number of individual endmember abundances that were assigned non-zero values when they were known to be zero. This is clinically relevant since, for example, PpIX abundance is used as a marker for malignancy. Having a false positive value could lead to unnecessary and harmful resection.

The same was repeated with all the human data, but the sparsity was evaluated instead of abundance error and false positive rate, which require ground truth values. For this, the sparsity of a vector was defined as the L0 norm, the number of non-zero elements in the vector. The sparsity of an unmixing was thus taken to be the mean L0 norm of all the computed abundance vectors. The false positive rate was defined as the number of individual endmember abundances that were assigned non-zero values when they should have been zero. The unmixing runtime per spectrum was measured as the total runtime for all 555666 spectra, divided by 555666. The runtime is very relevant for unmixing high-resolution data cubes in real time. The spectral angle error (SAM) measures the similarity of two spectra, akin to the cosine similarity[57]. The measured and reconstructed spectra are treated as high-dimensional vectors, and the cosine of the angle between them is computed using $\cos\theta = \frac{s_i \cdot (B\hat{c}_i)}{\|s_i\| \|B\hat{c}_i\|}$ where $s_i$ is the measured spectrum, and $B\hat{c}_i$ the fitted one. Since this compares the cosine of the angle, values close to 1 mean the spectra are similar. The abundance error is the mean square error (MSE) between the computed and expected abundance vectors: $e_{ab} = \frac{1}{n}\sum_{i=1}^n \|\hat{c}_i - c_{0,i}\|^2$. Similarly, the reconstruction error is the MSE of the measured and fitted spectra: $e_{rc} = \frac{1}{n}\sum_{i=1}^n \|B\hat{c}_i - s_i\|^2$.

Finally, to assess the value of the 9 chosen endmember spectra in the spectral library, the best unmixing technique was chosen, which gave accurate and sparse results, and the distribution of each endmember abundance across all the human data was determined. The code and spectral library are available in the Supplementary Materials.

## Results

First, the human data was unmixed with the five endmember spectra from[8] and with all 9 endmember spectra using NNLS. Two examples with varying PpIX content are shown in Fig. 5. It is qualitatively apparent that the unmixing with 9 basis spectra assigns non-zero abundances to many endmembers without truly improving the fit quality.

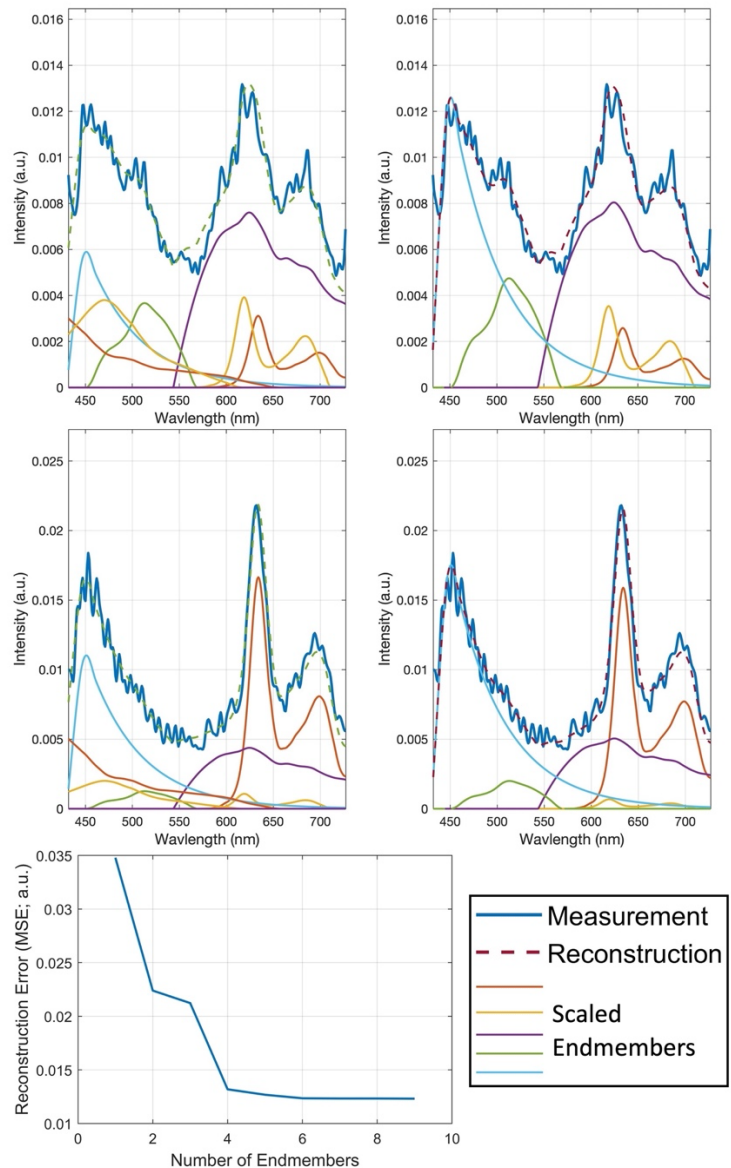

*Figure 5.* Two typical unmixings using NNLS with 9 spectra (left) and 5 spectra (right). We see that with 9 spectra, many of the endmembers are given non-zero abundances without visibly improving the quality of the fit, though the MSE does decrease. Below is the reconstruction error with increasing number of endmembers. PpIX$_{634}$, PpIX$_{620}$, lipofuscin, collagen, NADH, melanin, elastin, flavin, and FAD were added in that order. After 4 basis spectra, the error no longer improves significantly, but the additional spectra are often assigned non-zero abundance nonetheless.

The same result is shown looking at the reconstruction error in Fig. 5, where adding more spectra beyond 4 does not significantly improve performance despite the complexity of the reconstructions increasing. This suggests that it should be possible to simultaneously represent the human spectra with only 4 endmembers, using a sparse reconstruction algorithm.

**Data Distribution**

From the noise distributions of the measured data, the variance was obtained at every wavelength and plotted versus the average magnitude of the spectra at each wavelength. The result is shown in Fig. 6. The variance is linearly related to the mean, with a coefficient of determination of $R^2 = 0.81$ and a correlation coefficient of 0.90. The slope is 1.37, showing that the mean and variance are approximately equal. The reason why most of the points are near zero mean is that the PpIX spectrum is near zero for most wavelengths. This relation between the mean and variance suggests a Poisson distribution.

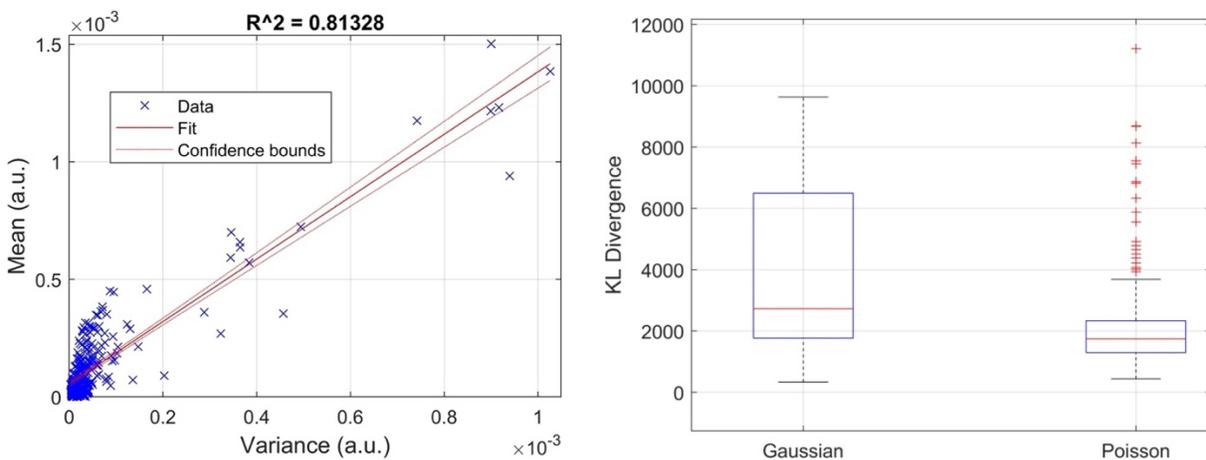

***Figure 6.*** *(Left) Signal mean versus noise variance of measured hyperspectral human data. The two are approximately linearly related with a slope of 1.37. (Right) Kullback-Leibler divergence of the measured data with modeled normal and Poisson distributions. The data is clearly much closer to the Poisson model.*

Next, the Kullback-Leibler divergence of the data distributions was analyzed at each wavelength with the two model distributions, Gaussian and Poisson, as described in the Methods section. The data distribution is far more nearly Poisson distributed than Gaussian, as shown in Fig. 6. The mean KL divergence for the Poisson distributions is 2120, while for the Gaussians, it is 3850, an increase of 82%. This is the rationale for pursuing a Poisson regression-based unmixing.

**Unmixing Tests**

As described in the Methods section, tests were performed on the six different unmixing algorithms, once with human data and once with simulated data. The runtimes of the methods are shown in Table 2. These depended on the regularization parameter, $\lambda$. After NNLS, which is a highly optimized MATLAB built-in function, SNPR is the fastest. The rate of ISTA and SNNLS are similar, while SLPR is an order of magnitude slower.

| Algorithm | Regularization ($\lambda$) | Runtime per spectrum (ms) |
|---|---|---|
| ISTA | 1.2 | 0.3828 |
|  | 1.4 | 0.179 |
|  | 1.6 | 0.196 |
| SNNLS | 0.3 | 0.2347 |
|  | 0.4 | 0.173 |
| SNPR | 0.25 | <u>0.162</u> |
|  | 0.35 | 0.2311 |
|  | 0.45 | 0.269 |
| NNLS | - | **0.143** |
| SLPR | 1 | 4.8 |
|  | 0.1 | 4.3 |
|  | 0.01 | 4.4 |

**Table 2.** *Runtimes of unmixing algorithms at relevant values of $\lambda$. The fastest is bolded, and the second best is underlined.*

The results on the simulated Data (1000 spectra) are shown in Table 3. SNPR equals the reconstruction error of NNLS, which is optimal in terms of MSE. Furthermore, SNPR with $\lambda = 0.35$ is second-best regarding false positives and abundance error. However, ISTA far outperforms all other methods in terms of false positives and clearly computes the most accurate abundance vectors. We see that despite its abundances matching the ground truth much better than any other algorithm, which is ultimately all that matters, the reconstruction error of ISTA is relatively large. This shows that reconstruction error is a flawed metric for the unmixing performance.

The number of false positives appears relatively large for all the algorithms except ISTA, but many false positive values are very small. Furthermore, there were far fewer false positive $PpIX_{634}$ values. For example, using ISTA with $\lambda = 1.4$, there are only 27 false positive $PpIX_{634}$ values or 0.27%. This is relevant because PpIX is often used as a malignant tissue marker.

| | | Simulated Data | | |
|---|---|---|---|---|
| Algorithm | Regularization ($\lambda$) | Reconstruction MSE $\times 10^{-4}$ | False Positives | Abundance MSE $\times 10^{-2}$ |
| ISTA | 1.2 | 6.34 ± 5.17 | 767 | 6.17 ± 9.07 |
|  | 1.4 | 8.09 ± 6.00 | 697 | **6.01 ± 0.86** |
|  | 1.6 | 11.0 ± 7.55 | **687** | 6.07 ± 8.12 |
| SNNLS | 0.3 | 3.54 ± 4.66 | 2081 | 7.87 ± 10.78 |
|  | 0.4 | 30.0 ± 51.0 | 1790 | 14.17 ± 16.09 |
| SNPR | 0.25 | **3.52 ± 4.66** | 2143 | 7.90 ± 10.81 |
|  | 0.35 | 3.79 ± 5.25 | <u>1621</u> | <u>7.81 ± 10.76</u> |

|  | 0.45 | 25.0 ± 29.0 | 2064 | 8.44 ± 16.51 |
| --- | --- | --- | --- | --- |
| NNLSQ | - | 3.52 ± 4.64 | 2222 | 8.05 ± 10.89 |
| SLPR | 1 | 4.16 ± 5.12 | 2664 | 8.42 ± 10.14 |
|  | 0.1 | 4.23 ± 5.14 | 2574 | 8.64 ± 10.32 |
|  | 0.01 | 4.24 ± 5.14 | 2568 | 8.67 ± 10.35 |
| Human Data | | | | |
| Algorithm | Regularization ($\lambda$) | Reconstruction MSE $\times 10^{-2}$ | Abundance L0 Norm | Reconstruction SAM |
| NNLS | (5 Spectra) | 12.7 ± 5.60 | 4.50 | 0.216 ± 0.105 |
|  | (9 Spectra) | 5.54 ± 9.60 | 5.53 | 0.120 ± 0.044 |
| SLPR | 1.5 | 5.94 ± 10.2 | 7.61 | <u>0.128 ± 0.053</u> |
| SNPR | 0.35 | **5.55 ± 9.60** | <u>5.25</u> | **0.120 ± 0.044** |
| ISTA | 1.4 | 6.60 ± 9.19 | **3.89** | 0.175 ± 0.089 |

*Table 3.* Unmixing performance on 1000 simulated spectra with ground truth abundance values and human data. The false positive column is the total number of abundances (out of 9000) that were incorrectly assigned nonzero values. Values are mean ± standard deviation when appropriate. The best values are bolded, and the second best are underlined. NNLS is the baseline and MSE-optimal, so it is not considered.

The results on human data are also shown in Table 3. It is not possible to evaluate abundance accuracy or false positive rate as there is no ground truth data. Instead, SAM and L0 norm are used. NNLS achieves the best reconstruction error, but again, SNPR is extremely close to optimal. Furthermore, SNPR gives the best SAM and second-highest sparsity. Again, ISTA achieves the best sparsity, at approximately 4 endmembers per measured spectrum, without greatly sacrificing the reconstruction error. The results from ISTA are, in fact, sparser than when only the 5 endmembers from Black et al.[8] are used, despite the reconstruction error also being half. The reconstruction error and SAM of SLPR are low, but this is likely due to substantial overfitting, with, on average, 7.6 of the 9 endmembers used in every unmixing.

It should be noted that a difference of 0.28 in the average L0 norm (Between SNPR and NNLS) equates to 155,586 fewer endmembers assigned non-zero values, or more than 1 in 4 spectra having one fewer endmember assigned. Thus, it is a substantial difference.

**Endmember Library**
Using ISTA with $\lambda = 1.4$, which gave the best results on the simulated spectra and the sparsest human data unmixing, the human data was again unmixed, and the distributions of the endmembers were analyzed, as shown in Table 4. PpIX$_{634}$, PpIX$_{620}$, lipofuscin, and collagen are by far the most common, followed by NADH and melanin. Elastin and flavins are sometimes needed, and FAD rarely. Overall, however, all spectra are used relatively frequently. In particular, the mean, median, and standard deviation values are shown in Table 4. Clearly, all spectra are important. Some are frequently present and in high abundance; others, like NADH, are less frequently present but important when they are, and still, others are often there but only in small quantities. However, all are essential for a complete and accurate dataset description.

|         | PpIX$_{634}$ | PpIX$_{620}$ | Lipof. | Flavin | NADH | FAD   | Collagen | Elastin | Melanin |
|---------|--------------|--------------|--------|--------|------|-------|----------|---------|---------|
| Mean    | 0.422        | 0.099        | 0.234  | 0.011  | 0.010| 0.001 | 0.186    | 0.018   | 0.019   |
| Median  | 0.477        | 0.063        | 0.122  | 0      | 0    | 0     | 0.128    | 0       | 0       |
| Std Dev | 0.328        | 0.105        | 0.251  | 0.027  | 0.055| 0.008 | 0.143    | 0.062   | 0.050   |
| %       | 84.1         | 66.3         | 70.1   | 21.9   | 7.3  | 16.5  | 96.4     | 14.0    | 29.0    |

*Table 4.* Statistics of fractional endmember abundances in human data using ISTA unmixing with regularization $\lambda = 1.4$. The last row indicates the percentage of the measurements that contain non-zero amounts of that endmember.

## Discussion

This paper has shown, using a large and diverse dataset of human brain tumor HSI images, that the measurements in HSI for fluorescence-guided resection of brain tumors are (1) Poisson distributed, (2) sparse in terms of endmember abundances, and (3) diverse in their fluorophore content. The consequence of the third point is that while each individual spectrum contains, on average, around 4 fluorophores to describe a whole dataset, a broader library of endmembers is needed. Therefore, this article has, for the first time in neurosurgery, compiled a library of 9 endmember spectra and paired them with sparse unmixing algorithms to represent the diversity of brain tumors without overfitting individual spectra. To do so, a simulation algorithm for human brain tumor HSI measurements was developed. A novel, simplified sparse Poisson regression method was also implemented, and five algorithms were tested, both on human and simulated data.

Overall, the SNPR and ISTA algorithms substantially outperformed all others. The outstanding performance of ISTA was a surprise as it ultimately solved the same optimization as SNNLS. However, it produced by far the most accurate endmember abundance vectors, while all other methods overfit to varying degrees. The second-best method in this respect was SNPR, which was slightly faster and produced better reconstructions, both in terms of SAM and L2 norm. Both results applied equally to simulated and human data. Therefore, depending on the application, SNPR ($\lambda = 0.35$) or ISTA ($\lambda = 1.4$) should be used: the former if reconstruction and/or speed are paramount, and the latter if endmember abundances and sparsity are more important.

This is not a detailed mathematical exploration of the unmixing problem in the context of neurosurgery, and various aspects, such as convergence rates or theoretical accuracy limits given by the Cramér-Rao lower bound of the estimator[58], are yet to be determined. In addition, further algorithms could be tested, for example ones making use of the Hessian, which is known, or using alternating direction method of multipliers (ADMM)[59]. Alternatively, FISTA is an improved version of ISTA, also from Beck and Teboulle[56]. There are many other adaptations of this algorithm, for example by Wei et al. and those reviewed in their introduction[60]. FISTA is shown in the Supplementary Material and was tested but performed much worse than all other methods and was thus excluded. This family of algorithms, as well as others, deserve a closer look.

Future work should also explore deep learning for spectral unmixing in brain tumor surgery. Much research in deep learning for general HSI unmixing exists and was reviewed briefly in the

Introduction section. However, neurosurgery has its own particular challenges. Additionally, it is crucial for this application and others that the unmixing results are explainable. Deep learning provides no guarantees for the output endmember abundances, which makes it difficult to trust them to guide the resection of brain tumors. Hence, classical methods with mathematical guarantees of optimality under closely studied conditions, as described in this paper, are valuable. Future work in explainable AI for HSI unmixing in neurosurgery may, however, improve performance.

With the spectral library and better unmixing method, it will also be interesting to see if the performance of machine learning classifiers that use the abundances improves. Previous work used 5 endmember abundances to classify tumor type, tumor margins, IDH mutation, and WHO grade with a relatively high degree of accuracy[12]. With 9 endmembers and more accurate values, these results will likely improve. Similarly, future work should revisit analyses of what endmembers are present in higher quantities or in different ratios in what types of tissues[3,9], to see if more concrete results can be discovered. Through this future work, this paper can gain direct clinical relevance. In addition, the improved accuracy, decreased false-positive PpIX abundances, and increased computational efficiency of the unmixing are all very useful for intraoperative guidance.

## Conclusion

This paper has shown that HSI fluorescence measurements of human brain data are Poisson distributed and have sparse abundance vectors, and are thus partial to sparse Poisson regression techniques for unmixing. In particular, a maximum likelihood algorithm was derived based on projected gradient descent with heavy-ball momentum and was shown to be both faster and more accurate for spectral unmixing than previous methods. Furthermore, an ISTA-based least squares algorithm outperformed all others in enforcing sparsity and accurately determining the fluorophore abundances underlying the noisy signal. It was also shown that, while sparse, the fluorophore content is diverse, so the presented spectral library of 9 fluorophores is essential. Together, the unmixing algorithms and spectral library compiled in this article will hopefully enable more accurate analysis of hyperspectral brain tumor data for intraoperative fluorescence guidance.

## Data Availability

The patient data used in this study is not available publicly. However, the endmember spectra are available in the supplementary material, as is MATLAB code for the unmixing algorithms and for creating the simulated spectra.

## Supplementary Material

Please see appendices, code, and endmember spectra at  github.com/dgblack/hsibrain

## Acknowledgments

We would like to thank Carl Zeiss Meditec AG (Oberkochen, Germany) for providing us with the OPMI pico system and the BLUE 400 filter.

## Author Contributions
*Conception and Design: D.B., E.S.M.*
*Acquisition of data: S.Ka., E.S.M.*
*Statistical Analysis and Interpretation: D.B., B.L., E.S.M.,*
*Drafting the article. D.B.*
*Critically revising the article: all authors*
*Technical support: A.D.I.*
*Study supervision: E.S.M.*

## References
1. Stepp, H. & Stummer, W. 5-ALA in the management of malignant glioma. *Lasers Surg Med* **50**, 399–419 (2018).
2. Stummer, W. *et al.* Fluorescence-guided surgery with 5-aminolevulinic acid for resection of malignant glioma: a randomised controlled multicentre phase III trial. *Lancet Oncology* **7**, 392–401 (2006).
3. Suero Molina, E. *et al.* Unraveling the blue shift in porphyrin fluorescence in glioma: The 620 nm peak and its potential significance in tumor biology. *Front Neurosci* **17**, 1261679 (2023).
4. Suero Molina, E., Kaneko, S., Black, D. & Stummer, W. 5-Aminolevulinic acid-induced porphyrin contents in various brain tumors: implications regarding imaging device design and their validation. *Neurosurgery* **89**, 1132–1140 (2021).
5. Schupper, A. J. *et al.* Fluorescence-Guided Surgery: A Review on Timing and Use in Brain Tumor Surgery. *Front Neurol* **12**, 682151 (2021).
6. Suero Molina, E., Kaneko, S., Black, D. & Stummer, W. 5-Aminolevulinic acid-induced porphyrin contents in various brain tumors: implications regarding imaging device design and their validation. *Neurosurgery* **89**, 1132–1140 (2021).
7. Iordache, M.-D., Bioucas-Dias, J. M. & Plaza, A. Sparse unmixing of hyperspectral data. *IEEE Transactions on Geoscience and Remote Sensing* **49**, 2014–2039 (2011).
8. Black, D. *et al.* Characterization of autofluorescence and quantitative protoporphyrin IX biomarkers for optical spectroscopy-guided glioma surgery. *Scientific Reports 2021 11:1* **11**, 1–12 (2021).
9. Alston, L. *et al.* Spectral complexity of 5-ALA induced PpIX fluorescence in guided surgery: a clinical study towards the discrimination of healthy tissue and margin boundaries in high and low grade gliomas. *Biomed Opt Express* **10**, 2478–2492 (2019).
10. Fürtjes, G. *et al.* Intraoperative microscopic autofluorescence detection and characterization in brain tumors using stimulated Raman histology and two-photon fluorescence. *Front Oncol* **13**, 1146031 (2023).
11. Jarry, G., Henry, F. & Kaiser, R. Anisotropy and multiple scattering in thick mammalian tissues. *JOSA A* **17**, 149–153 (2000).
12. Black, D. *et al.* Towards Machine Learning-based Quantitative Hyperspectral Image Guidance for Brain Tumor Resection. (2023).
13. Walke, A. *et al.* Challenges in, and recommendations for, hyperspectral imaging in ex vivo malignant glioma biopsy measurements. *Sci Rep* **13**, 3829 (2023).

14. Leclerc, P. *et al.* Machine learning-based prediction of glioma margin from 5-ALA induced PpIX fluorescence spectroscopy. *Sci Rep* **10**, 1462 (2020).
15. Kaneko, S. *et al.* Fluorescence real-time kinetics of protoporphyrin IX after 5-ALA administration in low-grade glioma. *J Neurosurg* **1**, 1–7 (2021).
16. Suero Molina, E., Black, D., Kaneko, S., Müther, M. & Stummer, W. Double dose of 5-aminolevulinic acid and its effect on protoporphyrin IX accumulation in low-grade glioma. *J Neurosurg* **137**, 943–952 (2022).
17. Martinez, B. *et al.* Most Relevant Spectral Bands Identification for Brain Cancer Detection Using Hyperspectral Imaging. *Sensors 2019, Vol. 19, Page 5481* **19**, 5481 (2019).
18. Baig, N. *et al.* Empirical Mode Decomposition Based Hyperspectral Data Analysis for Brain Tumor Classification. *Proceedings of the Annual International Conference of the IEEE Engineering in Medicine and Biology Society, EMBS* 2274–2277 (2021) doi:10.1109/EMBC46164.2021.9629676.
19. Giannantonio, T. *et al.* Intra-operative brain tumor detection with deep learning-optimized hyperspectral imaging. *https://doi.org/10.1117/12.2646999* **12373**, 80–98 (2023).
20. Bioucas-Dias, J. M. *et al.* Hyperspectral unmixing overview: Geometrical, statistical, and sparse regression-based approaches. *IEEE J Sel Top Appl Earth Obs Remote Sens* **5**, 354–379 (2012).
21. Nielsen, A. A. Spectral mixture analysis: Linear and semi-parametric full and iterated partial unmixing in multi-and hyperspectral image data. *J Math Imaging Vis* **15**, 17–37 (2001).
22. Heylen, R., Burazerovic, D. & Scheunders, P. Fully constrained least squares spectral unmixing by simplex projection. *IEEE Transactions on Geoscience and Remote Sensing* **49**, 4112–4122 (2011).
23. Bhatt, J. S. & Joshi, M. V. Deep learning in hyperspectral unmixing: A review. in *IGARSS 2020-2020 IEEE International Geoscience and Remote Sensing Symposium* 2189–2192 (2020).
24. Wang, C. J., Li, H. & Tang, Y. Y. Hyperspectral Unmixing Using Deep Learning. *International Conference on Wavelet Analysis and Pattern Recognition* **2019-July**, (2019).
25. Zhang, X., Sun, Y., Zhang, J., Wu, P. & Jiao, L. Hyperspectral unmixing via deep convolutional neural networks. *IEEE Geoscience and Remote Sensing Letters* **15**, 1755–1759 (2018).
26. Hong, D. *et al.* Endmember-guided unmixing network (EGU-Net): A general deep learning framework for self-supervised hyperspectral unmixing. *IEEE Trans Neural Netw Learn Syst* **33**, 6518–6531 (2021).
27. Qu, Y. & Qi, H. uDAS: An untied denoising autoencoder with sparsity for spectral unmixing. *IEEE Transactions on Geoscience and Remote Sensing* **57**, 1698–1712 (2018).
28. Licciardi, G. A. & Del Frate, F. Pixel unmixing in hyperspectral data by means of neural networks. *IEEE Transactions on Geoscience and Remote Sensing* **49**, 4163–4172 (2011).
29. Quintano, C., Fernández-Manso, A., Shimabukuro, Y. E. & Pereira, G. Spectral unmixing. *Int J Remote Sens* **33**, 5307–5340 (2012).
30. Heylen, R., Parente, M. & Gader, P. A review of nonlinear hyperspectral unmixing methods. *IEEE J Sel Top Appl Earth Obs Remote Sens* **7**, 1844–1868 (2014).

31. Heylen, R. & Scheunders, P. A multilinear mixing model for nonlinear spectral unmixing. *IEEE transactions on geoscience and remote sensing* **54**, 240–251 (2015).
32. Chen, J., Richard, C., Ferrari, A. & Honeine, P. Nonlinear unmixing of hyperspectral data with partially linear least-squares support vector regression. in *2013 IEEE International Conference on Acoustics, Speech and Signal Processing* 2174–2178 (2013).
33. Geladi, P. & Kowalski, B. R. Partial Least-Squares Regression - a Tutorial. *Anal Chim Acta* **185**, 1–17 (1986).
34. Bro, R. & DeJong, S. A fast non-negativity-constrained least squares algorithm. *J Chemom* **11**, 393–401 (1997).
35. Coates, P. B. Photomultiplier noise statistics. *J Phys D Appl Phys* **5**, 915 (1972).
36. Wang, R. *et al.* Unmixing biological fluorescence image data with sparse and low-rank Poisson regression. *Bioinformatics* **39**, btad159 (2023).
37. Rajabi, R. & Ghassemian, H. Spectral unmixing of hyperspectral imagery using multilayer NMF. *IEEE Geoscience and Remote Sensing Letters* **12**, 38–42 (2014).
38. Khan, U., Paheding, S., Elkin, C. P. & Devabhaktuni, V. K. Trends in Deep Learning for Medical Hyperspectral Image Analysis. *IEEE Access* **9**, 79534–79548 (2021).
39. Cui, R. *et al.* Deep Learning in Medical Hyperspectral Images: A Review. *Sensors 2022, Vol. 22, Page 9790* **22**, 9790 (2022).
40. Jia, S. *et al.* A survey: Deep learning for hyperspectral image classification with few labeled samples. *Neurocomputing* **448**, 179–204 (2021).
41. Ebner, M. *et al.* Intraoperative hyperspectral label-free imaging: from system design to first-in-patient translation. vol. 54 294003 Preprint at https://doi.org/10.1088/1361-6463/ABFBF6 (2021).
42. Ruiz, L. *et al.* Multiclass Brain Tumor Classification Using Hyperspectral Imaging and Supervised Machine Learning. *2020 35th Conference on Design of Circuits and Integrated Systems, DCIS 2020* (2020) doi:10.1109/DCIS51330.2020.9268650.
43. Urbanos, G. *et al.* Supervised Machine Learning Methods and Hyperspectral Imaging Techniques Jointly Applied for Brain Cancer Classification. *Sensors 2021, Vol. 21, Page 3827* **21**, 3827 (2021).
44. Fabelo, H. *et al.* Deep Learning-Based Framework for In Vivo Identification of Glioblastoma Tumor using Hyperspectral Images of Human Brain. vol. 19 920 Preprint at https://doi.org/10.3390/S19040920 (2019).
45. Leon, R. *et al.* Hyperspectral imaging benchmark based on machine learning for intraoperative brain tumour detection. *npj Precision Oncology 2023 7:1* **7**, 1–17 (2023).
46. Fabelo, H. *et al.* Spatio-spectral classification of hyperspectral images for brain cancer detection during surgical operations. *PLoS One* **13**, e0193721 (2018).
47. Rinesh, S. *et al.* Investigations on Brain Tumor Classification Using Hybrid Machine Learning Algorithms. *J Healthc Eng* **2022**, (2022).
48. Fabelo, H. *et al.* In-vivo hyperspectral human brain image database for brain cancer detection. *IEEE Access* **7**, 39098–39116 (2019).
49. Hao, Q. *et al.* Fusing Multiple Deep Models for in Vivo Human Brain Hyperspectral Image Classification to Identify Glioblastoma Tumor. *IEEE Trans Instrum Meas* **70**, (2021).

50. Valdés, P. A. *et al.* A spectrally constrained dual-band normalization technique for protoporphyrin IX quantification in fluorescence-guided surgery. *Opt Lett* **37**, 1817–1819 (2012).
51. Valdés, P. A. *et al.* Quantitative, spectrally-resolved intraoperative fluorescence imaging. *Sci Rep* **2**, 798 (2012).
52. Berkson, J. Estimation by least squares and by maximum likelihood. in *Proceedings of the Third Berkeley Symposium on Mathematical Statistics and Probability, Volume 1: Contributions to the Theory of Statistics* vol. 3 1–12 (1956).
53. Shlens, J. Notes on Kullback-Leibler Divergence and Likelihood. (2014).
54. Lawson, C. L. & Hanson, R. J. *Solving Least Squares Problems*. (SIAM, 1995).
55. Polyak, B. T. Some methods of speeding up the convergence of iteration methods. *USSR Computational Mathematics and Mathematical Physics* **4**, 1–17 (1964).
56. Beck, A. & Teboulle, M. A fast iterative shrinkage-thresholding algorithm for linear inverse problems. *SIAM J Imaging Sci* **2**, 183–202 (2009).
57. Yuhas, R. H., Goetz, A. F. H. & Boardman, J. W. Discrimination among semi-arid landscape endmembers using the Spectral Angle Mapper (SAM) algorithm. *JPL, Summaries of the Third Annual JPL Airborne Geoscience Workshop. Volume 1: AVIRIS Workshop* (1992).
58. Hodges, J. L. & Lehmann, E. L. Some applications of the Cramer-Rao inequality. in *Proceedings of the Second Berkeley Symposium on Mathematical Statistics and Probability* vol. 2 13–23 (1951).
59. Boyd, S., Parikh, N., Chu, E., Peleato, B. & Eckstein, J. Distributed optimization and statistical learning via the alternating direction method of multipliers. *Foundations and Trends in Machine Learning* **3**, 1–122 (2010).
60. Wei, J. *et al.* A Faster and More Accurate Iterative Threshold Algorithm for Signal Reconstruction in Compressed Sensing. *Sensors (Basel)* **22**, (2022).